\documentclass[12pt,draftcls,journal,letterpaper,oneside,onecolumn]{IEEEtran}

\usepackage{cite}
\usepackage[pdftex]{graphicx}
\usepackage[cmex10]{amsmath}
\usepackage{amssymb}
\usepackage{amsfonts}

\usepackage{algorithm}
\usepackage{algorithmic}
\usepackage[caption=false,font=footnotesize]{subfig}

\usepackage{color}
\usepackage{array}
\usepackage{mdwmath}
\usepackage{mdwtab}
\usepackage{eqparbox}

\date{}

\begin{document}
\title{Big Spectrum Data: The New Resource for Cognitive Wireless Networking}

\author{\IEEEauthorblockN{Guoru Ding, Qihui~Wu, Jinlong Wang,\\ PLA University of Science and Technology, China\\
Yu-Dong Yao, Stevens Institute of Technology, USA
}
\thanks{This work is supported by the National Natural Science Foundation of China (Grant No. 61172062 and No. 61301160), Jiangsu Province Natural Science Foundation (Grant No. BK2011116), and in part by the National Basic Research Program of China (Grant No. 2009CB320400).}}
\maketitle

\begin{abstract}
The era of Big Data is here now, which has brought both unprecedented opportunities and critical challenges. In this article, from a perspective of cognitive wireless networking, we start with a definition of Big Spectrum Data by analyzing its characteristics in terms of six V's, i.e., volume, variety, velocity, veracity, viability, and value. We then present a high-level tutorial on research frontiers in Big Spectrum Data analytics to guide the development of practical algorithms. We also highlight Big Spectrum Data as the new resource for cognitive wireless networking by presenting the emerging use cases.
\end{abstract}

%
\IEEEpeerreviewmaketitle

\IEEEPARstart{W}e now live in an era of data deluge. We live in a world that has more than a billion transistors per human; a world with more than 4 billion mobile phone subscribers and about 30 billion radio frequency identification tags produced globally within the last two years~\cite{Understanding-Big-Data}. All these sensors generate data. Sadly, much of this data is simply thrown away, because of the lack of efficient mechanisms to derive value from it. This fact motivates the worldwide increasing interests in Big Data.

Although there are still debates on whether Big Data is a big opportunity or a big bubble, Big Data is here now and is going to transform how we gain insights and how we make decisions in the future~\cite{Viktor}. To leverage the Big Data opportunities and challenges, many governments, organizations and academic institutions come forward to take initiates. For example, the US government has announced a Big Data Research and Development (R$\&$D) initiative in March 2012, to develop and improve the tools and techniques needed to access, organize, and analyze Big Data and to use Big Data for scientific research and national security~\cite{White-house}. The China government has supported a growing number of national grants/programs on Big Data R$\&$D in the past couple of years~\cite{China-NSF}. Many famous transnational corporations, such as Google, Yahoo!, Microsoft, IBM, Intel, and Huawei have also begun to develop various business use cases of Big Data, such as enhanced 360-degree view of the customer, security and intelligence extension, smart city and internet of things (IoT), etc.


In this article, we investigate Big Data from a new perspective: Cognitive wireless networking. The motivation is driven by the following facts. The rapid development of mobile internet, mobile social networking and wireless IoT are all greatly hampered by limited radio spectrum resource. However, a number of spectrum measurement campaigns~(see e.g.,~\cite{Spectrum_prediction,Spectrum_models}) have demonstrated that radio spectrum is not physically scarce, but was vastly underutilized, with temporal and geographical variations in the utilization raging from 15$\%$ to 85$\%$. To improve radio spectrum utilization, cognitive wireless networking~\cite{Haykin_2005} is known as a promising paradigm, by mainly providing radio environment awareness-based dynamic spectrum sharing among heterogeneous networks~\cite{SPMag2013}. To obtain radio environment awareness, Geolocation Spectrum Database~\cite{Senseless_2012} and Radio Environment Map~\cite{REM-MassiveData} are well recognized as the key enablers, which depend highly on efficient mechanisms to derive value from massive and complex spectrum data sets. However, to the authors' best knowledge, so far there is no study to systematically tackle this vital topic.


Motivated by the observations above, this article propose to exploit \emph{Big Spectrum Data} as the new resource for cognitive wireless networking to improve radio spectrum utilization. First we analyze the key characteristics of Big Spectrum Data in detail. We then present a high-level tutorial on research frontiers in Big Spectrum Data analytics. We also present the emerging use cases.
\\

\textbf{\emph{Big Spectrum Data: Key Characteristics}}

\textbf{\emph{What is Spectrum Data and What is Big Spectrum Data?}}

Spectrum Data refers to all the data that are related to radio environment awareness, mainly including:
\begin{itemize}
  \item Radio spectrum state data (idle or busy, signal energy levels, signal features, etc) in the time, space and frequency dimensions.
  \item User or device data, e.g., device ID, device capability, user spectrum requirement and user feedback, etc.
  \item Environment side information, e.g., terrain data, meteorologic and hydrographic data, etc.
\end{itemize}

\begin{figure}[!b]
\centering
\includegraphics[width=0.6\linewidth]{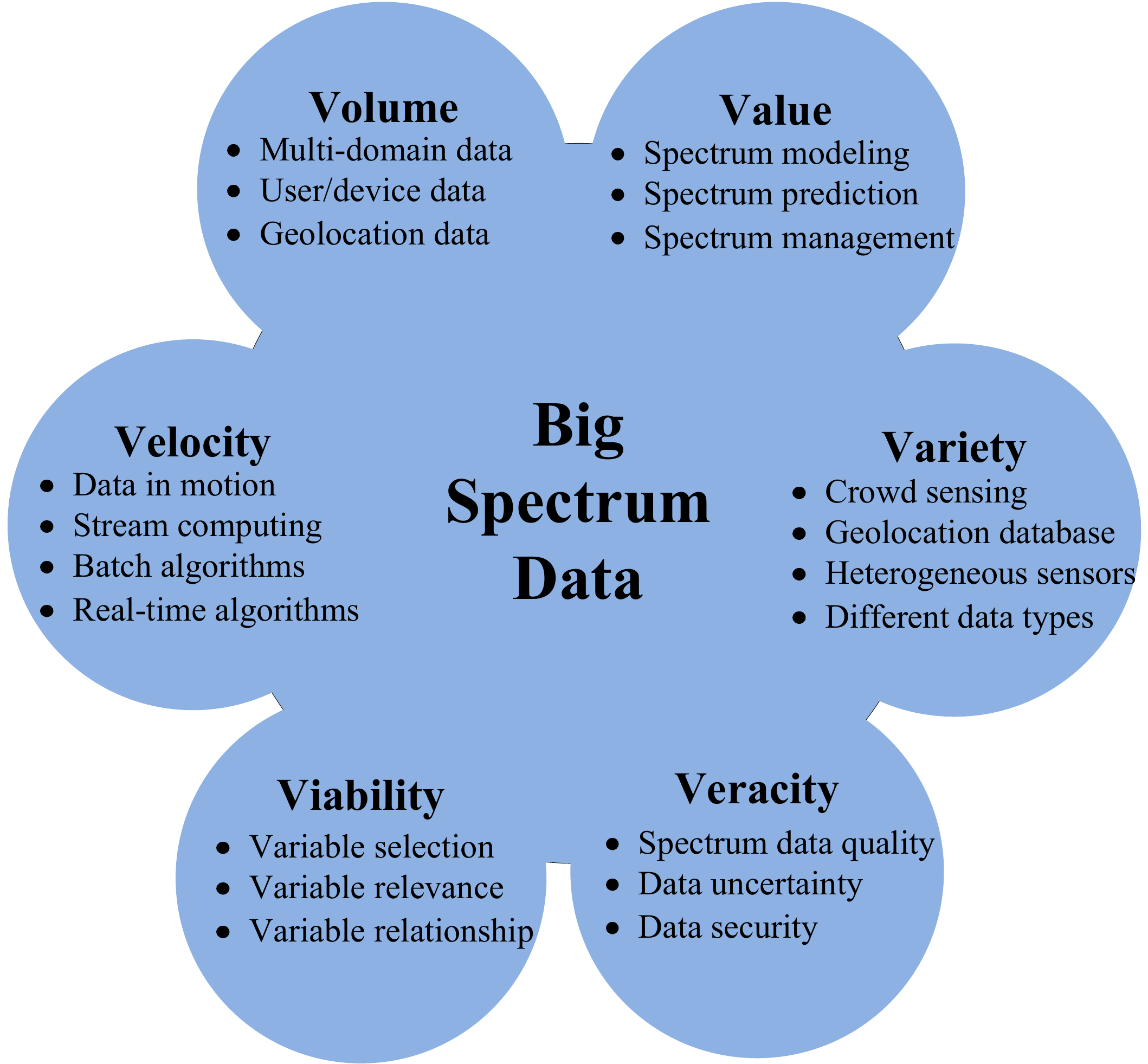}
\caption{Key characteristics of Big Spectrum Data.}
\label{Fig-Big-Spectrum-Data}
\end{figure}

Big Spectrum Data, a specific pattern of Big Data in wireless or radio domain, in short, refers to massive and complex spectrum data that can't be processed or analyzed using traditional systems and tools. The term ``Big Spectrum Data" is a bit of a misnomer since it implies that the only characteristic is its sheer size (size is one of them, but there are often more). Specifically, as shown in Fig. \ref{Fig-Big-Spectrum-Data}, Big Spectrum Data can be characterized by six keywords: \emph{volume}, \emph{variety}, \emph{velocity}, \emph{veracity}, \emph{viability} and \emph{value}. The first four keywords are learned from what the IBM refers to as (general) ``Big Data"~\cite{IBM-real-world-use}. We absorb this knowledge for the understanding of Big Spectrum Data here and extend it with two other vital characteristics, \emph{viability} and \emph{value}. In the following, we explicitly define each characteristic in detail.
\\
\\

\textbf{\emph{The Volume of Big Spectrum Data}}

\begin{figure}[!t]
\centering
\includegraphics[width=0.6\linewidth]{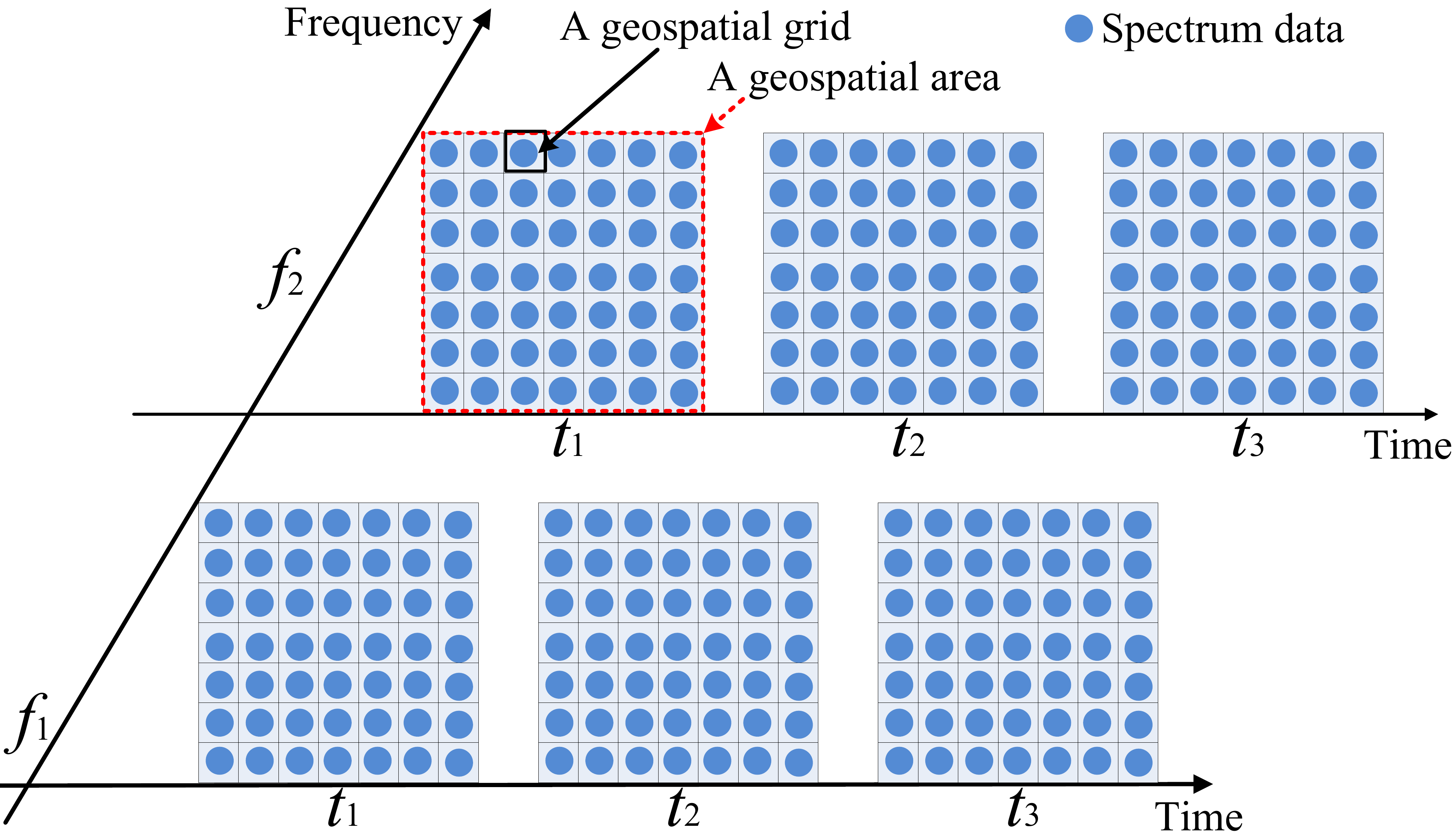}
\caption{Spectrum data in time-space-frequency multi-dimensional space.}
\label{Fig-Spatial-Temporal-Frequency}
\end{figure}

Perhaps volume is the characteristic most associated with Big Spectrum Data, which refers to the mass quantities of spectrum data that are used to improve decision-making for better radio spectrum utilization. The sheer volume of spectrum data being stored and processed is exploding at an unprecedented rate, which is mainly driven by the need to gain a full understanding of radio spectrum dynamics. Taking only the radio spectrum state data as an instance, as shown in Fig. \ref{Fig-Spatial-Temporal-Frequency}, if we use 1 Byte to represent the spectrum data (e.g., the signal energy level) in a geospatial grid of 100 m $\!\times\!$ 100 m, a resolution frequency band of 100 kHz, and a time slot of 100 ms, after a time duration of one week, the total data size, in a frequency band ranging from 0 to 5 GHz and a geospatial area of 100 Km $\!\times\!$ 100 Km, can be as large as:
\begin{align}
&7~\rm{days/week} \!\times\! 24~\rm{hours/day} \!\times\! 3600~\rm{seconds/hour}\nonumber\\
&~~~\times\! \frac{1 ~\rm{s}}{100~\rm{ms}} \!\times\! \frac{5~\rm{GHz}}{100~\rm{kHz}} \!\times\! \frac{100~\rm{Km} \!\times\! 100~\rm{Km}}{100~\rm{m} \!\times\! 100~\rm{m}} \!\times\! 1~ \rm{Byte}\nonumber\\
&= 3.024 \!\times\! 10^{17}~\rm{Byte}\nonumber\\
&= 3.024 \!\times\! 10^5 ~\rm{Terabyte ~(TB)} \nonumber\\
&= 3.024 \!\times\! 10^2~ \rm{Petabyte ~(PB)}.
\end{align}

The volume of spectrum state data grows with the time duration, the frequency range, and the spatial scale of interest, as well as the corresponding resolution in each dimension. Moreover, if we further take into consideration the user or device data, the terrain data, and the meteorologic and hydrographic data, the volume of spectrum data will become much bigger and may change from petabytes to zettabytes ($\rm{ZBs},~1~\rm{ZB} = 10^{21}$ Byte). Notably, there is no need to provide a strict definition on how big is big, and the focus should be on how to derive value from the spectrum data of sheer volume that can't be processed well in traditional systems and tools. Now, a natural question arises: What's the value of spectrum data in such a big scale and who will care about it?
\\

\textbf{\emph{The Value of Big Spectrum Data}}

Briefly, Big Spectrum Data allows a more complete picture of radio spectrum usage and deeper understanding of the hidden patterns behind spectrum state evolution and spectrum utilization. Specifically, the value of Big Spectrum Data can be embodied at least in the following three aspects:
\begin{itemize}
  \item \emph{Comprehensive spectrum modeling}, which depends highly on extensive spectrum measurement data~\cite{Spectrum_models}, reflects spectrum usage patterns in the time, space and frequency dimensions, and serves as the cornerstone for the analysis, design and simulation of cognitive wireless networks.
  \item \emph{Accurate spectrum prediction}, which essentially infers unknown spectrum state from known spectrum data, by effectively exploiting the statistical correlations extracted from Big Spectrum Data. Accurate spectrum prediction can enable efficient spectrum usage by looking into the future. Just as Amazon has used huge records of consumers' historical behaviors to successfully predict their future preferences, accurate spectrum prediction is also possible since many real world spectrum measurements~(see e.g.,~\cite{Spectrum_prediction}) have revealed that radio spectrum usage is not completely random, actually, correlations exists across time slots, frequency bands, and locations.
  \item \emph{Flexible spectrum management}. It is known that the current radio spectrum scarcity problem actually results from static and inflexible spectrum licensing and allocation policies, rather than the physical scarcity of usable radio frequencies. With Big Spectrum Data, both real-time and statistical information on spectrum usage in a much finer granularity can be obtained, which makes spectrum management much more flexible by performing dynamic spectrum assignment and smart interference management to improve spectrum utilization.
\end{itemize}

The value can be shared by many people. Spectrum regulators (such as FCC and Ofcom) can use it to establish flexible spectrum policies. Telecommunication operators can have more usable radio frequencies. Companies, such as Spectrum Bridge Inc. and Google Inc., can provide new services on spectrum database construction and maintenance. Last but not least, consumers can enjoy mobile internet, mobile social networking and wireless IoT, with little worry about the limited radio spectrum resource.
\\

\textbf{\emph{The Variety of Big Spectrum Data}}

Variety means that spectrum data generally come from various sources and are of different types. Nowadays, to enable radio environment awareness, spectrum data mainly have two sources: i) spectrum sensing and ii) geolocation spectrum database. In spectrum sensing approaches, spectrum data are obtained from a crowd of (heterogeneous) spectrum sensors, either specialized spectrum measurement equipments or personal mobile devices. These data are fused to determine the spectrum state (idle or busy) via hypotheses testing~\cite{SPMag2013}. In spectrum database approaches, a centralized database provides location-based spectrum services for each user, predicting the spectrum state at that user's location by using a combination of sophisticated radio signal propagation modeling, GPS localization data, high-resolution terrain data, and the parameters (e.g., transmission power and antenna height) of working transmitters~\cite{Senseless_2012}.

Spectrum data are also of heterogeneous types. Besides the traditional structured data suitable for database systems, the emergence of semi-structured and unstructured spectrum data create new challenges. For example, spectrum data are generally labeled with geo-spatial and temporal lags, creating challenges in maintaining coherence across spatial scales and time. Spectrum data sometimes involve networks and graphs, with inferential questions hinging on semantically rich notions such as ``centrality" and ``influence."
\\




\textbf{\emph{The Veracity of Big Spectrum Data}}

Veracity refers to quality of spectrum data. We include veracity as the fourth key characteristic of Big Spectrum Data to emphasize the importance of addressing and managing the \emph{uncertainty} on spectrum data quality.

Some spectrum data are inherently uncertain resulted from, for example, ubiquitous noise, fading, and shadowing in wireless environment; localization errors due to the GPS sensors' bouncing among the skyscrapers of Manhattan; imperfect terrain data; and ever-changing weather conditions, etc.

Other spectrum data are uncertain due to the involvement of the sentiment and truthfulness in humans. For example: lack of efficient incentive mechanisms for data contributors may result in spectrum data incomplete; lack of secure defense schemes to malicious contributors may result in spectrum data corrupted.


Humans, by nature, dislike uncertainty, but just ignoring it can create even more problems than the uncertainty itself. To address the uncertainty, executives will need to acknowledge it, embrace it and determine how to use it to their advantage.

Data analysts need to create context around Big Spectrum Data. The key idea is that no spectrum data exists in isolation, and there are correlations between each data and its neighbors in time, frequency, and space dimensions. By properly exploiting these correlations, either explicitly or implicitly, the uncertainty can be reduced and the quality of spectrum data can be improved. One way to achieve this is through data fusion, where combining multiple spectrum data from less reliable sources creates a more accurate and useful data. Another way is through data filtering, where abnormal spectrum data are filtered out by comparing the differences or divergences among Big Spectrum Data. Other ways include data cleansing, data completion, and data recovery, etc.
\\






\textbf{\emph{The Viability of Big Spectrum Data}}

Viability guides the selection of the attributes and factors that are most likely to predict outcomes that matter most to the value of Big Spectrum Data. For instance, to build a radio propagation predictive model from Big Spectrum Data, we're not simply collecting a large number of records, actually, we're collecting multidimensional data that spans a broadening array of variables. With so many varieties of data and variables, we want to quickly and cost-effectively test and confirm a particular variable's relevance on the model we are building, before investing a fully featured and over-sophisticated model.

The secret is to uncover the latent, hidden relationships among these variables, which can begin with simple hypotheses such as,
  \begin{itemize}
    \item Does the weather (meteorologic and hydrographic) condition affect radio signal propagation?
    \item Is 2D geolcation good enough for spectrum modeling? or Is 3D terrain data further needed?
    \item How do time, frequency, geolocation and other factors all converge to predict spectrum state evolution?
  \end{itemize}

Validation of each hypothesis is needed before we take further action. The viability of Big Spectrum Data guides to rethink the fundamental tradeoff between comprehensiveness or accurateness and complexity of the model we plan to build.
\\


\textbf{\emph{The Velocity of Big Spectrum Data}}

Velocity impacts latency$-$the lag time between when spectrum data are captured and when decisions based on them are made. Today, spectrum data is continually being generated at such a speed that it is more and more difficult for traditional systems to handle in time. Some delay-tolerant tasks, such as statistical spectrum modeling of radio spectrum usage, can possibly be performed in an off-line manner using batch algorithms. However, for time-sensitive processes such as spectrum prediction, the shelf life of spectrum data is short, which must be analyzed in real-time to be of value for decision-making, in which case online or incremental processing techniques on Big Spectrum Data are urgently needed.

As shown in Fig. \ref{Fig-Spatial-Temporal-Frequency}, if we treat the given geospatial area as an image, with each spectrum data corresponding to a pixel, when the spectrum state evolves with time, we can obtain a 3D video: space$-$2D and frequency$-$1D. Consequently, we should accommodate the velocity of Big Spectrum Data in a new way named \emph{spectrum data in motion}: The speed at which the spectrum data is flowing or streaming. To tackle this challenge, the key idea is to leverage massively parallel processing techniques to analyze data while it is streaming, so we can understand what is happening in real time, make better decisions, and improve spectrum utilization. According to~\cite{Understanding-Big-Data}, IBM InfoSphere Streams platform with the capability of Big Data scale stream computing may be a good choice.



In short, effectively deriving value from Big Spectrum Data requires that one performs analytics against the volume, variety, veracity and viability of data, while it is still in motion, not just after it is at rest.
\\




\begin{figure}[!t]
\centering
\includegraphics[width=0.6\linewidth]{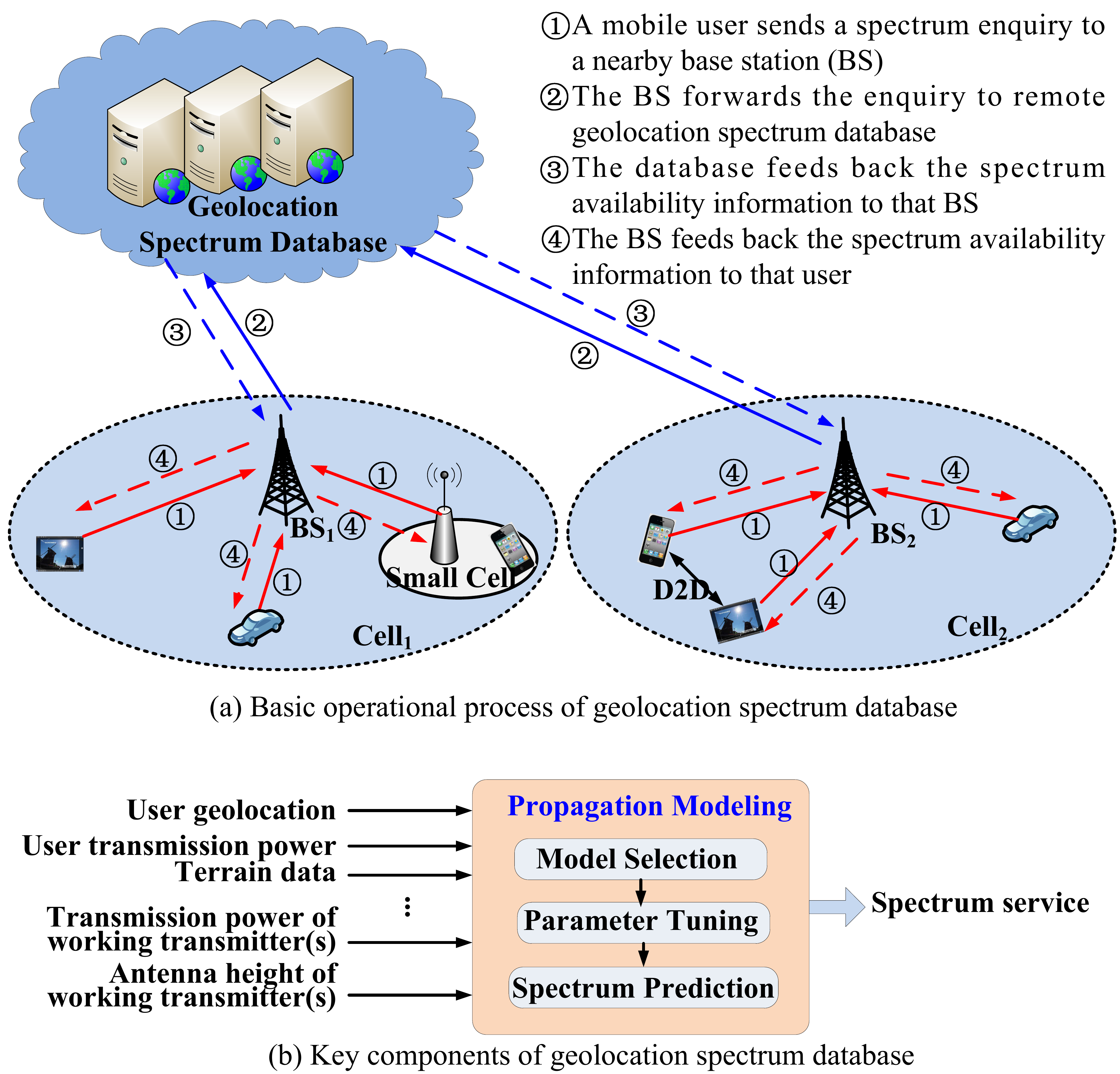}
\caption{Geolocation spectrum database-assisted dynamic spectrum sharing.}
\label{Visio-use-case}
\end{figure}

\textbf{\emph{Emerging Use Cases}}

So far, the emerging use cases mainly focus on exploiting Big Spectrum Data to improve radio spectrum utilization by providing radio environment awareness through geolocation spectrum database (see, e.g.,~\cite{Senseless_2012}), which can provide location-based spectrum services for device-to-device (D2D) communications, femtocells, small cells, and general mobile users with localization capability.

Fig. \ref{Visio-use-case}(a) shows the basic operational process. Firstly, a mobile user with spectrum requirement sends an enquiry (embedded with its geolocation and transmission power) to the nearby base station (BS). Then, the BS forwards this enquiry to a remote geolocation spectrum database. The database calculates the set of vacant frequency band(s) at the location of that user using a combination of radio signal propagation models, terrain data, and up-to-date parameters of the working transmitters (see Fig. \ref{Visio-use-case}(b)), and then feeds back the spectrum availability information to that user via the BS.

As shown in Fig. \ref{Visio-use-case}(b), one key component of the geolocation spectrum database is the selection of proper propagation models. Recent spectrum measurement campaigns (see, e.g.,~\cite{survey_PathLoss}) have demonstrated that current propagation models are suitable for nationwide radio coverage planning, but perform poorly at predicting accurate path loss even in relatively simple outdoor environments. Moreover, the suitability of a particular propagation model and the corresponding parameter setup are of great discrepancies in different environments.

A promising direction is to combine the idea of mobile crowd sensing into geolocation spectrum database~\cite{JSAC-D2D}. Personal devices such as smartphones, tablets, and in-vehicle sensors can be employed to collect massive spectrum measurements or Big Spectrum Data, which can be further used to calibrate the propagation models and improve the accuracy of spectrum prediction. Another similar application is to develop Radio Environment Map~\cite{REM-MassiveData}, visualizing Big Spectrum Data to assist decision-making of spectrum regulators and telecommunication operators. Many other applications, such as spectrum database-assisted spectrum selection for opportunistic networks, can also be further developed, the success of which depends highly on efficient mechanisms and algorithms as discussed below.
\\

\textbf{\emph{Frontiers in Big Spectrum Data Analytics}}

Big Spectrum Data brings both unprecedented opportunities and critical challenges, which require new methods of data analytics, to acquire knowledge on radio spectrum usage from spectrum data and improve decision-making for better spectrum utilization. In the following, we present a brief tutorial on the research frontiers.
\\
%
%

%
%
%

\textbf{\emph{Volume: Parallel and Distributed Computing}}

Increasing the volume of spectrum data on a given problem (e.g., spectrum modeling) generally increases the capability (e.g., statistical power) to address the problem. However, very-large-scale spectrum data introduce critical data management challenges. The most notable one is harnessing the computational power required to analyze the big data.

The clear trend to improve the computational power is to make increasing use of parallel and distributed computing. According to~\cite{Frontiers}, the recent directions include hardware parallelism, multi-core CPUs, flash memory, data stream management systems, MapReduce, Hadoop, cloud systems, parallel and distributed databases, and parallel programming languages and systems, etc. Currently, the main limitations include that parallel and distributed softwares are difficult to write, and parallel and distributed algorithms and systems are difficult to configure and maintain. Consequently, achieving greater use of the power of parallel and distributed computing requires further innovations that simplify their use and maintenance.
\\

\textbf{\emph{Velocity: Time-Aware Data and Real-Time Analysis}}

The acquisition and prediction of real-time radio spectrum state requires spectrum data in motion, which forces spectrum database to store massive time-aware spectrum data, and provide real-time analytics to minimize the latency between when spectrum data are captured and when decisions based on them are made. Currently, two major challenges include:
\begin{itemize}
  \item Distributed real-time acquisition, storage, and transmission of massive time-aware spectrum data among sources.
  \item Mixed online and batch algorithms for spectrum data processing, representation, and inference. A typical method is to build two separate and loosely coupled systems. A online streaming system might provide real-time alerting, while historical (statistical) analyses are made on a batch-oriented system.
      \\
\end{itemize}

\textbf{\emph{Variety: Hybrid Computer/Human Data Analysis}}

Conventionally, spectrum data are collected by expensive specialized equipments in various measurement campaigns~(see e.g.,~\cite{Spectrum_prediction,Spectrum_models}), and then processed by powerful computers. Recently, the convergence of sensing, communication, and computational power on personal mobile devices (such as cellular phones and in-vehicle sensors) creates an unprecedented opportunity for the building of crowdsourced spectrum database~\cite{JSAC-D2D}, which harnesses human activity to contribute spectrum data and accomplish large-scale spectrum measurements.

Perhaps more interesting development is to leverage human intelligence to further explicitly involve in data-analysis process, since it is well known that computers and people excel at very different types of tasks. Although the fields of artificial intelligence and machine learning have made great progress in recent years, people's ability to disambiguate, understand context, and make subjective judgments, exceed the capabilities of even the most sophisticated computing systems. Therefore, hybrid computer and human data analysis is a promising R$\&$D direction.
\\

\textbf{\emph{Viability: Resources, Limitations and Tradeoffs}}

Big Spectrum Data is a consequence of complex systems with many kinds of variables, and viability guides the selection of the most suitable one(s) by considering various resources, limitations, and tradeoffs as a whole. Big Spectrum Data analytics involve many types of resources: 1) storage resource; 2) computational resource; 3) communication resource; and 4) energy, etc. Resources are generally limited or constrained, and resource-efficient designs (such as green computing) are of great value in the future.

Fundamental performance limits and tradeoffs involved in Big Spectrum Data analytics are also of particular interest. For example, how much spectrum data and resources are needed in order to build a spectrum model with a given accuracy and confidence? What are the lower bounds on the amount of energy or time required to collect and process spectrum data with a large-scale? Which are the dominating factors or variables that affect the comprehensiveness-complexity tradeoff in spectrum modeling process?
\\

%
%
%
%
%
%
%
%
%
%
%
%
%
%
%
%



\textbf{\emph{Veracity: Uncertainty, Integrity and Security}}

Veracity relates to the efforts to improve the quality of spectrum data, which are affected by a number of factors, notably uncertainty, integrity and security.

Although each of the factors outlined above have been studied in previous work for small spectrum data sets, new challenges or puzzles arise when spectrum data become massive. For example, while sampling error generally decreases with increasing sample size, bias does not$-$big data does not help overcome bad bias~\cite{Frontiers}. Moreover, are missing data less of an issue, because we have so much data that we can afford to lose some measurements? Can we simply discard spectrum measurements with missing entries? Because Big Spectrum Data are exposed to many sources of contamination, sometimes through malicious intervention~\cite{ShuiYu}, can models be built that self-protect against there various sources? Can the dirty or abnormal data get washed out in massive spectrum data analysis? While the answers are probably ``no" in general, but it may well be that procedures that were deemed inefficient for small data might be reasonable with massive data.


%
%
%

The research frontiers discussed above are not in isolation, but are coupled in practical applications of Big Spectrum Data. Moreover, we have just provided a general profile of the challenges and opportunities. The more complete picture is open for future studies.
\\

\textbf{\emph{Conclusions and Discussions}}

This article proposed to exploit Big Spectrum Data as the new resource for cognitive wireless networking to improve radio spectrum utilization. Both opportunities and challenges are discussed by first analyzing six-V-characteristics of Big Spectrum Data, and then providing a brief tutorial on the research frontiers as well as the emerging use cases.

The R$\&$D necessary for Big Spectrum Data goes well beyond the province of any single discipline, and there is the need for a thoroughly interdisciplinary enterprise. Solutions to the challenges ahead will require ideas from computer science and statistics, with essential contributions also needed from applied mathematics, from optimization theory, and from various engineering areas, notably wireless networking, signal processing and information theory.
\\

\end{document}